\documentclass[runningheads]{llncs}
\usepackage{graphicx}
\usepackage{amsmath}
\usepackage{array}
\usepackage{floatrow}
\usepackage{ tabularx}
\usepackage{ makecell}
\usepackage{graphicx,stackengine}

\usepackage{caption}
\usepackage{subcaption}
\usepackage{xcolor}
\usepackage{listings}
\usepackage{todonotes}

\begin{document}
\title{
REPTILE: A Proactive Real-Time Deep Reinforcement Learning
Self-adaptive Framework}

\author{Flavio Corradini\inst{1}\orcidID{0000-1111-2222-3333} \and
Michele Loreti\inst{1}\orcidID{1111-2222-3333-4444} \and
Marco Piangerelli\inst{1}\orcidID{2222--3333-4444-5555} \and 
Giacomo Rocchetti\inst{1}}

\authorrunning{F. Corradini et al.}
%
\institute{University of Camerino, Camerino MC 62031, Italy \\
\email{\{flavio.corradini, michele.loreti, marco.piangerelli\}@unicam.it\\ giacomo.rocchetti@studenti.unicam.it}\\
}
%
\maketitle              
\begin{abstract}

In this work a general framework is proposed to support the
development of software systems that are able to adapt their behaviour according to
the operating environment changes. The proposed approach, named
REPTILE, works in a complete proactive manner and relies on Deep
Reinforcement Learning-based agents to react to events, referred
as novelties, that can affect the expected behaviour of the system. In
our framework, two types of novelties are taken into account: those related to
the context/environment and those related to the physical architecture
itself. The framework, predicting those novelties before their occurrence,
extracts time-changing models of the environment and uses a suitable
Markov Decision Process to deal with the real-time setting. Moreover,
the architecture of our RL agent evolves based on the possible actions
that can be taken.

\end{abstract}
\section{Intoduction}
\label{intro}

The last decades witnessed a significant evolution in software designing enactment as the process of managing, developing and maintaining software is becoming more and more complex.
The always increasing expansion of Internet together with the advent of a plethora of \emph{smart devices}, posed new challenges to engineers and developers that have to design and maintain software infrastructure that must be able to \emph{adapt their behaviour} to new requirements and updates in the system architecture often unknown at design time.
These can have a great impact on the Quality of Service (QoS) offered by the system, thus impacting the final user experience.

In this context, Kephart and Chess stated that ``\emph{the only option remaining is autonomic computing systems that can manage themselves given high-level objectives from administrators}''~\cite{kephart2003vision}. Such systems are commonly called \textit{Self-Adaptive Systems} (SASs) and they have been extensively researched and developed in the last twenty years. 
The main feature of SASs is to be able to operate in an environment that is changing in an unpredictable way without any, or limited, external control. 
For this reason, a big effort has made to design methodologies and techniques supporting SASs adaptation.

Several approaches have been proposed to design and develop SASs~\cite{krupitzer2015survey}. 
In particular we can mention here the \emph{model-based}, the \emph{architecture-based}, the \emph{control theory-based}, the \emph{formal modeling and verification-approach}  and the \textit{Learning-based} approaches. 
The \emph{model-based} approach uses model-driven engineering, filling incomplete information at design time with the usage of \emph{run-time models} (MUSIC framework \cite{hallsteinsen2012development}).
The \emph{architecture-based} approach use the concept of the high level structure of software for different activities. One of the most well-known frameworks that implements this approach is the \emph{Rainbow framework}, which provides an architecture layer whose components define adaptation plans~\cite{garlan2004rainbow}.
In the \emph{control theory-based} approach the system continuously takes measurements (from sensors) and performs adjustments to keep the measured variable in a defined range in a process called \emph{feedback loop}~\cite{muller2008visibility}. 
Together with the implementation of SASs, there is the need to prove the correctness of such systems using \emph{formal modeling and verification} approaches. A common method is the \emph{formal reference model for self-adaptation} (FORMS), which uses the Z notation in order to define the whole system, from its interaction with the environment to the adaptation~\cite{weyns2010forms}.
The latter approach, the \textit{learning-based} one, given the increasing popularity of the field of \emph{Machine Learning} (ML), mainly in the last decade, affected a lot the research of Self-Adaptive Systems using ML algorithms for a variety of purposes: from the use of genetic algorithms for the run-time adaptation of mobile applications in~\cite{pascual2013run}, or the use of \emph{Reinforcement Learning} (RL) for an architecture-based adaptation~\cite{kim2009reinforcement}, to the application of \emph{Deep Learning} (DL) for SASs in network security~\cite{papamartzivanos2019introducing,maimo2018self}.

Moreover, challenges such as the analysis of \textit{streams of data in real-time} and the extraction of environmental models in order to aid a proactive adaptability for unforeseen changes have to be adequately taken into account. To this extent, we can mention a framework for self adaptive systems proposed in~\cite{elkhodary2010fusion}. The authors tackle the problem of unforeseen changes in the environment by learning the impact of adaptation decisions on the system's goal applying a reactive, i.e. post-hoc or after-the-fact, model.
A more refined framework is the one proposed in \cite{klos2015adaptive}. It is based on the MAPE-K framework, extended with a simplified RL component.

Finally, the IoTArchML framework refers to a proactive SAS applied to  \emph{iot} systems~\cite{muccini2019machine}. In particular, it is designed for the control of system's QoS parameters, using a proactive approach to anticipate eventual changes before a QoS deviation event.

In this work we propose a general \emph{framework for software SASs dealing with real-time streams of data}.
Our framework works in a complete proactive manner.
It uses of \emph{Deep Reinforcement Learning} (DRL) algorithms, in order to adapt itself to eventual \emph{novelties}, i.e. new events that can be managed in order to work properly. In our framework, as already pointed out in~\cite{klos2015adaptive}, we identify two types of novelties: those relating to the \textit{context/environment} and those relating to the \textit{physical architecture} itself.
The framework, predicting those novelties before their occurrence, extract time-changing models of the environment and uses a modified Markov Decision Process to deal with the real-time setting as shown in~\cite{NEURIPSramstedt2019real}.
Moreover the architecture of our RL agent evolves based on the possible actions that can be taken.

The paper is structured as follows: in section~\ref{sec:final_framework} the self-adaptability framework is described, section~\ref{sec:Adaptation_Process} presents the explanation of the adaptability mechanisms, in section~\ref{sec:results} we show the results obtained by applying the framework to a real case study and, finally, in section~\ref{sec:conclusions} we discuss the results.

\section{REPTILE Architecture}
\label{sec:final_framework}

\begin{figure}[tbp]
    \centering
     \includegraphics[height=0.8\linewidth]{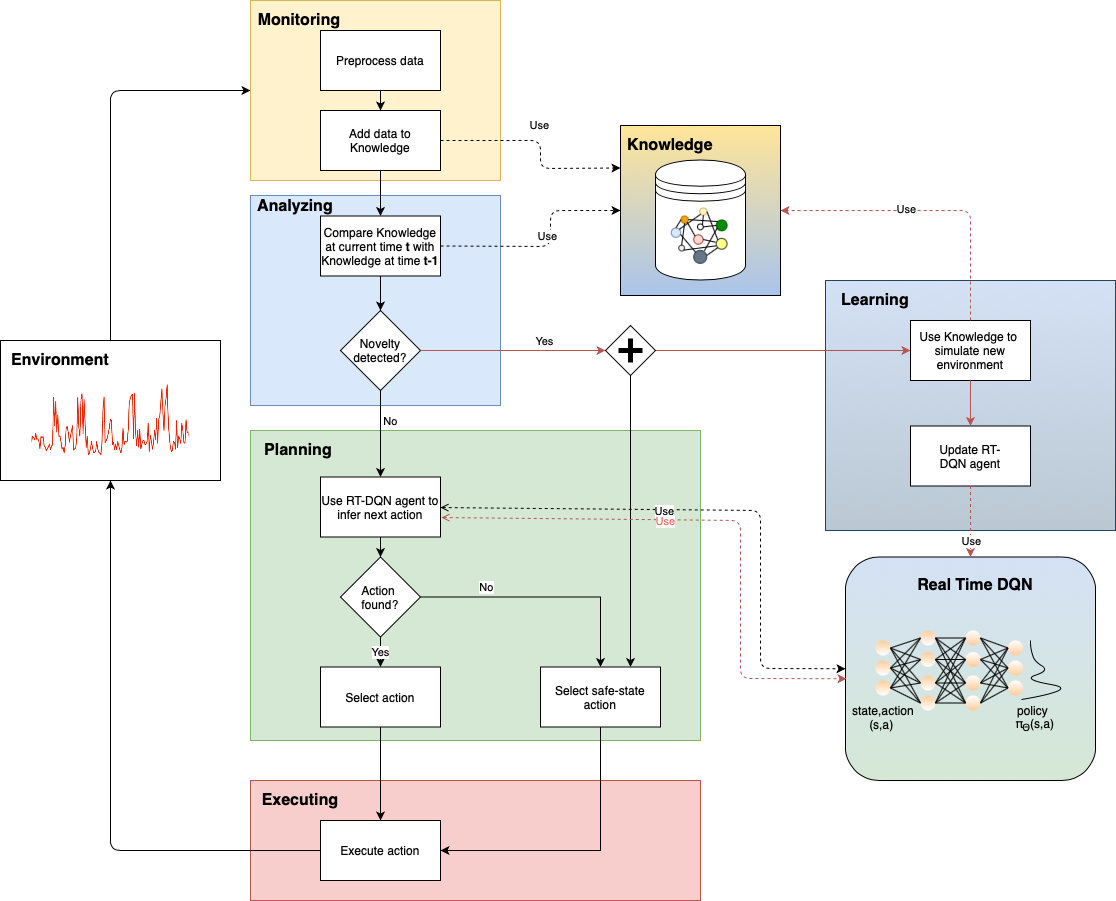}
    \caption{Overview of REPTILE.}
    \label{fig:final_framework}
\end{figure}

In this Section we provide a general overview of our framework, REPTILE. It is based on the well known IBM's MAPE-K architecture enriched with a \emph{Reinforcement Deep Learning module}.
The whole architecture of REPTILE is depicted in 
Figure~\ref{fig:final_framework}. It consists of six modules: the five from the standard MAPE-K architecture (\emph{Monitoring}, \emph{Analyzing}, \emph{Planning}, \emph{Executing}, and \emph{Knowledge}) plus a \emph{Reinforcement Deep Learning} one.
Below, the role of these modules is outlined together with their use in a simple running scenario.  

\noindent
\textbf{Monitoring (yellow)}. It continuously collects data from the environment and performs an extraction of useful features of the system, based on the domain application. It then sends data to the \textit{Knowledge} component.

\begin{example}[Running Case Study (1/6)]
\label{run:1}
To better illustrate the technicalities of our framework we present here a running case study. 
We applied our framework to a Heating, Ventilation and Air Conditioning (HVAC) system. The reason behind the choice of this particular application domain is its simplicity. A simulator for the environment was implemented in Python: it computes the next internal  temperature  of  a  single  room based on the actions performed by the system. The simulator is made up by the following three devices:
\begin{itemize}
\item \textbf{Heater}: it produces power (Watts) in order to heat the room.
\item \textbf{Cooler}: it produces power (Watts) in order to cool the room.
\item \textbf{Window}: managed by a binary signal (open or closed).
\end{itemize}

The gained/lost temperature inside a room is obtained by calculating the heat loss between the room and the outside, through walls and windows, and converting the power produced by the heater as a heat gain and the one produced by the cooler as heat loss. 

Both the description of the room and the available devices are customizable. A device is any instrument that can modify the internal temperature with its action. Table~\ref{tab:simulator_parameters} shows the simulations parameters. External temperatures data have been pre-processes from two existing datasets \cite{beach_dataset,beniaguev_2017}.

\end{example}

\noindent \emph{\textbf{}} 

\begin{table}[]
    \centering
    \begin{tabular}{ |c|c| }
        \hline
         \textbf{Parameter} & \textbf{Value} \\
        \hline
        Room Width & 5 \\
        Room Height & 3 \\
        Room Depth & 3 \\
        Wall R-Value & 3.6 \\
        Window R-Value & 0.176 \\
        Air Changes per Hour & 1.7 \\
        N. Heaters & 1 \\
        N. Coolers & 1 \\
        N. Windows & 1 \\
        Heater Power (W) & 0, 200, 400 \\
        Cooler Power (W) & 0, 400 \\
        \hline
    \end{tabular}
    \caption{Room parameters used in simulations}
    \label{tab:simulator_parameters}
\end{table}

\noindent
\textbf{Knowledge (yellow and blue).} It receives data from the \textit{Monitoring} module and it builds two types of models of the environment: \emph{time-varying models},  $\mathcal{M}_{t,{\bar{\alpha}}}^{\mathcal{TV}}$. and \emph{global 
models}, $\mathcal{M}_{\xi}^{\mathcal{G}}$. The former is used for monitoring the model in the short-period, while the latter represents the comprehensive knowledge of the environment. In the previous notation $t$ represents a generic time step, $\bar{\alpha}$ the vector containing the parameters of the model. Superscript $\mathcal{TV}$ indicates that we are modeling the time-varying, while $\mathcal{G}$ stands for Global, where $\xi = 1,2,3...,k$ is the index identifying each model .
Moreover, this module contains some pre-trained agents, $\mathcal{M}_{\xi}^{\mathcal{A}}$, where $\xi = 1,2,3...,n$ is the index identifying each model and the superscript $\mathcal{A}$ stands for agent.%

\begin{example}[Running Case Study (2/6)]
\label{run:2}
In our running case study, the \emph{knowledge} consists in a set of \emph{models} ( $\mathcal{M}_{t,{\bar{\alpha}}}^{\mathcal{TV}}$, $\mathcal{M}_{\xi}^{\mathcal{G}}$, $\mathcal{M}_{\xi}^{\mathcal{A}}$). 
Each $\mathcal{M}_{t,{\bar{\alpha}}}^{\mathcal{TV}}$ is represented by using an \textbf{Auto-Regressive}, \textbf{Moving Average} and \textbf{Integrated} (ARIMA) model~\cite{box2015time}.

Such models are described formally by the following equation:
\begin{equation}
    y'_t = c + \epsilon_t + \sum_{i = 1}^p \varphi_i y'_{t-i} + \sum_{i=1}^q \theta_i \epsilon_{t-i}
\end{equation}

where $\theta_i$ and $\phi_i$ are the parameters of the model, $p$ and $q$ are the \textit{orders}, respectively of the auto-regression and the moving average contributions; 
The orders of an ARIMA model are defined as follows:
\begin{itemize}
\item $p$: the number of lag observations included in the model, also called the lag order.
\item $q$: the size of the moving average window, also called the order of moving average.
\end{itemize}
$\epsilon_t \overset{iid}{\sim} N(0, \sigma^2)$
and  $c = \mu + \varphi_0$ with $\mu$ the expected value of $y_t$ and  $y_t' = y_t-y_{t-1}$. To obtain the model we used a number of data points $d=30$.
Each time a new data point arrives, we compute a new model discarding the oldest point end incorporating the new one.
The $\mathcal{M}_{\xi}^{\mathcal{G}}$ models can be obtained using more sophisticated techniques. In this work, for simplicity, we already had such models. 
Finally, $\mathcal{M}_{\xi}^{\mathcal{A}}$ is the model obtained using some DRL-based model.
\end{example}

\noindent
\textbf{Analyzing (blue).} It inspects the data stored in the \textit{Knowledge}, that is a \emph{model} of the collected data, and uses it to forecast possible evolution. At each step this module is able to detect potential unseen behaviours, i.e. \textbf{novelties}.

We can distinguish two kinds of novelties: \textit{Contextual novelty} and \textit{Architectural novelty}~\cite{klos2015adaptive}. The former indicates a  \emph{novelties} are mainly inferred by comparing the predictions done at the previous steps with the one at the current step.  
For instance, the module can detect that an observable, namely the \emph{temperature}, is going to run out of the expected range in the next fifteen minutes, since the predicted value increased in the last observations. 
The latter refer to a change in the interaction between the system and the environment due to a new capability that is now available. For instance, a new device is installed that can affect the observations.

Actually, this module is also able to detect novelties in a \emph{reactive} manner, i.e. if the observable  violates a certain "soft" constraint for a given amount of time. For instance, whenever \emph{temperature} is in the Intermediate zone (see Section~\ref{subsec:RTRL}) for $t$ previous time steps.

\begin{example}[Running Case Study (3/6)] 
\label{run:3}
In this case study, we identify novelties in the following way:
    \begin{itemize}
    
    \item a new device has been inserted into the system (architectural novelty).
    \item the internal temperature does not stay in the interval [$18$ \textdegree{}C, $22$ \textdegree{}C] (contextual novelty).

    \end{itemize}

First of all, REPTILE looks for new devices or changes in the devices themselves, i.e. e new heater with different powers of use.
If such a component is detected, the system begins the adaptation procedure in order to adapt the structure of the \emph{learning} module to manage new actions.
If no new device is found the system continues to behave as usual.
Moreover, REPTILE uses also the predictions of the next three points for deciding about the onset of the adaptability process: if the third point (the farthest in time) is not in the comfort zone then the adaptability process is triggered, otherwise everything remains the same (\emph{proactive} process). In the meantime, the $T=12$ previous time steps are monitored in order to establish if system has been keeping violating constraints since 12 time steps: in that case the adaptability process is triggered, too (\emph{reactive} process).
The \emph{executing} module, i.e, the DRL-based agent performs an action every 15 minutes while the ARIMA model is derived every 5 minutes meaning that the system had 3 time steps to try to correct its behavior.
\end{example}

\noindent
\textbf{Learning (blue and grey).} It is the crucial component of the framework and it is activated when a novelty is foreseen. This module is responsible for the choice of the optimal adaptive policy.
It uses the \textit{Knowledge} and a DRL-based agent (the same used by the Planning module) for the adaptation process. Once a novelty is detected, the \textit{Learning} component is invoked, otherwise the \textit{Plan} component will decide the next action based on the current DRL-based agent.
Generally speaking, DRL is about maximizing a reward function, i.e. a cost function, in order to make the DRL-based agent to learn how to behave in an environment (refer to Subsection~\ref{subsec:RTRL} for the reward function used in this work). 

\begin{example}[Running Case Study (4/6)]
\label{run:4}
We used the Deep Q-Learning algorithm for implementing the \emph{Learning} module: in particular we implemented a Duelling Deep Q-network~\cite{wang2016dueling}. The Dueling Deep Q-network splits the single stream of fully connected layers in two separate streams, one for the \textit{value function, $V$,} and one for the \textit{advantage function, $A$}, defined as:
\begin{equation}
    A^\pi(s, a) = Q^\pi(s, a) - V^\pi(s)
    \label{eq:advantage_function}
\end{equation}

In our implementation of the Dueling Deep-Q network we used the \textit{real time markov decision process (RTMDP)} framework~\cite{NEURIPSramstedt2019real}, see Section~\ref{subsec:RTRL} for more details. Accordingly, states and actions evolve simultaneously. 
The state vector is then denoted as a state-action pair $\boldsymbol{x}_t = (s_t, a_t)$.
The actions in input to the Duelling Deep Q-network are encoded using a one-hot encoding of the combinations of the basic actions of the three devices.
For example, referring to Table~\ref{tab:reward_parameters}, there are $12$ actions (3 for the Heater, 2 for the Cooler and 2 for the Window). Using a possible one-hot encoding, Table~\ref{tab:actions} shows all the possible actions:
\begin{table}[!h]
    \centering
    \begin{tabular}{ |c|c|c|c|c| }
        \hline
         \textbf{Heater} & \textbf{Cooler} & \textbf{Window}& \textbf{$a_t$}\\
        \hline
        0    & 0   & CLOSE & 1,0,0,0,0,0,0,0,0,0,0,0\\
        0    & 0   & OPEN  & 0,1,0,0,0,0,0,0,0,0,0,0\\
        0    & 200 & CLOSE & 0,0,1,0,0,0,0,0,0,0,0,0\\
        0    & 200 & OPEN  & 0,0,0,1,0,0,0,0,0,0,0,0 \\
        200  & 0   & CLOSE & 0,0,0,0,1,0,0,0,0,0,0,0 \\
        200  & 0   & OPEN  & 0,0,0,0,0,1,0,0,0,0,0,0 \\
        200  & 200 & CLOSE & 0,0,0,0,0,0,1,0,0,0,0,0 \\
        200  & 200 & OPEN  & 0,0,0,0,0,0,0,1,0,0,0,0 \\
        400  & 0   & CLOSE & 0,0,0,0,0,0,0,0,1,0,0,0\\
        400  & 0   & OPEN  & 0,0,0,0,0,0,0,0,0,1,0,0 \\
        400  & 200 & CLOSE & 0,0,0,0,0,0,0,0,0,0,1,0\\
        400  & 200 & OPEN  & 0,0,0,0,0,0,0,0,0,0,0,1\\
        \hline
    \end{tabular}
    \caption{One-Hot encodind of the actions $a_t$}
    \label{tab:actions}
\end{table}
If the vector $a_{t}=(1,0,0,0,0,0,0,0,0,0,0,0)$ meaning Heater = 0 W, Cooler = 0 W, Window = CLOSE, and $s_{t}=(T_{ext} = 17, T_{in}=19)$, then $\textbf{x}_{t} = (17,19,1,0,0,0,0,0,0,0,0,0,0,0)$.
As shown in Figure~\ref{fig:networks_architecures}, our Duelling Deep-Q Networks is made by 3 fully connected hidden layers each of which contains 256 neurons. The Input layer contains 14 neurons and the Output ones consist of 12 neurons.
\begin{figure}[!h]
    \centering
    \includegraphics[angle=90,origin=c, width = 0.65\linewidth]{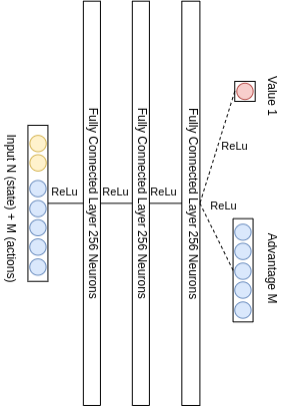}
    \caption[Deep Reinforcement Learning Architectures used]{The Dueling Deep Q-Network used in this work.}
    \label{fig:networks_architecures}
\end{figure}

\end{example}

\noindent
\textbf{Planning (green).} This component simply uses the DRL-based agent to choose the next action to perform.\\
\begin{example}[Running Case Study (5/6)]
\label{run:5}
Given in input $\textbf{x}_{t} = (17,19,1,0,0,0,0,\\0,0,0,0,0,0,0)$, then the Deep Q-Network chooses the next action to perform as  $a_{t+1}=(0,0,0,1,0,0,0,0,0,0,0,0)$ (Heater = 0 W, Cooler = 200 W, Window = OPEN).  
\end{example}

\noindent
\textbf{Executing (red).} Just as a usual MAPE-K, this component is in charge of executing the action planned by the \textit{Planning} component.\\
\begin{example}[Running Case Study (6/6)]
\label{run:6}
The Executor will closed the window and switch off the heater
    
\end{example}

\subsection{Real Time Reinforcement Learning Agent}
\label{subsec:RTRL}
The classical idea of a RL-based agent is the following: the agent receives the state from the environment, it selects an action according to a transition distribution and it performs that action, changing the environment. This scenario is modeled by using \emph{Markov Decision Process} (MDP)~\cite{sutton2018reinforcement}. 
The fundamental problem with this conception is that the environment is considered \textit{paused} during the action selection and that is why MDP is referred to \emph{Turn-based Markov Decision Process} (TBMDP) in ~\cite{NEURIPSramstedt2019real}. While it is suited for problems such as board games, it is certainly not suited for real-time applications
In order to derive the $\mathcal{M}_{\xi}^{\mathcal{A}}$, where $\xi = 1,2,3...,n$ is the index assign to each model and the superscript $\mathcal{A}$ indicates that we are speaking about a model of the Agent,
we implemented a Dueling Deep Q-network (Figure~\ref{fig:networks_architecures}), using the RTRL paradigm. In RL, the goal is maximizing the reward function to let the agent learn how to behave in a given environment. 
It has an experience replay and a target network, performing a soft update, to optimize the training phase. 

The exploration-exploitation trade-off is given by $\epsilon$: the chosen action at time step $t$, will be a greedy action (exploit) with probability (1-$\epsilon$) or may be a random action (explore) with probability of $\epsilon$.
Then it decrease the parameter $\epsilon$ by an $\epsilon$-decay value. The soft update is regulated by the parameter $\tau$. The list of parameters used in the study are shown in Table~\ref{tab:agent_params}.

\begin{table}[!h]
    \centering
    \begin{tabular}{ |c|c| }
        \hline
         \textbf{Parameter} & \textbf{Value} \\
        \hline
        Memory Size & 1000000 \\
        Batch Size & 256 \\
        $\epsilon$-decay & 0.0002 \\
        $\tau$ & 0.005 \\
        Hidden Layers Neurons & 256 \\ 
        Training Episodes & 600 \\
        \hline
    \end{tabular}
    \caption{Reinforcement Learning Agent parameters.}
    \label{tab:agent_params}
\end{table}


To  enable  the  agent  to  calculate  the  reward  of  its  actions  and  to  manage  the  training  process, the Open AI Gym toolkit has been used~\cite{brockman2016openai}.  
The advantage of using this tool is that it can be used by every type of algorithm and it does not need any knowledge about the RL agent interacting with it.  In fact, an Open AI environment only needs the action chosen by the agent in order to return the next state and the reward for that action.  
Every step of the environment is considered to be scheduled at 15 minutes intervals.

\subsubsection{Reward Function}
\label{sssec:reward}

A reward function balancing \textbf{energy consumption} and \textbf{thermal comfort} was chosen.
The two aspects are conflicting: in order to have an ideal thermal comfort, consumption of energy is needed, while saving energy could result in thermal discomfort. In this work, we modified the reward function defined in~\cite{brandi2020deep}. Let us define the \textbf{setpoint} as the desired comfort temperature set by the user and let us define two parameters $\epsilon, \delta > 0$ such that $\epsilon < \delta$. 
\begin{itemize}
    \item \textit{Comfort Zone}: area in the interval $[setpoint - \epsilon, setpoint + \epsilon]$
    \item \textit{Intermediate Zone}: two areas enclosed in the intervals $(setpoint - \delta, setpoint - \epsilon]$ and $(setpoint + \epsilon, setpoint + \delta]$.
    \item \textit{Danger Zone}: the two remaining intervals $(- \infty, setpoint - \delta)$, $(setpoint + \delta, + \infty)$.
\end{itemize}

The reward function is defined as follows:
\begin{equation}
    \begin{cases}
        0, & \text{if}\ setpoint - \epsilon \leq T_{in} \leq setpoint + \epsilon \\
        - \beta \cdot energy - \rho \cdot (setpoint - T_{in})^2, & \text{if}\ setpoint - \delta \leq T_{in} < setpoint - \epsilon \\
        & \lor\ setpoint + \epsilon < T_{in} \leq setpoint + \delta   \\
        - \beta \cdot energy - \rho \cdot (|setpoint - T_{in}|)^3, & \text{otherwise} \\
    \end{cases}
\end{equation}

where $T_{in}$ is the internal temperature, $\beta$ and $\rho$ are parameters that permit to balance the trade-off between the energy consumption and the thermal comfort. Table~\ref{tab:reward_parameters} shows the values of these parameters used in the simulations.

\begin{table}[!ht]
    \centering
    \begin{tabular}{ |c|c| }
        \hline
         \textbf{Parameter} & \textbf{Value} \\
        \hline
        Setpoint & 20 \\
        $\epsilon$ & 2 \\
        $\delta$ & 4 \\
        $\beta$ & 0.05 \\
        $\rho$ & 1 \\ 
        \hline
    \end{tabular}
    \caption{Reward function parameters used.}
    \label{tab:reward_parameters}
\end{table}

\section{Self-Adaptive strategy}
\label{sec:Adaptation_Process}

To managing the two different kinds of novelties, we implement three different self-adaptive strategies: the \textit{Model Switching Adaptation}, the \textit{Architectural Adaptation} and the \emph{DQN-Knowledge-based Adaptation}. The latter method always starts only if the first two fail.

\subsubsection{Model Switching Adaptation}
REPTILE applies this strategy when a \emph{context novelty} is detected. 
All the available pre-trained models $\mathcal{M}_{1}^{\mathcal{A}}$, $\mathcal{M}_{2}^{\mathcal{A}}$,...,$\mathcal{M}_{k}^{\mathcal{A}}$ are evaluated and the best one, in terms of some evaluation parameters, is selected. Our selecting criterion is based on returning the greatest reward using the current data.

\subsubsection{Architectural Adaptation}
This strategy of adaptation is based on the idea that the DRL-agent can learn to perform new actions if it is aware of having instruments allowing them: equipped with new devices there is a change in the number of potential actions the agent can perform. 
The solution we adopted is pretty straightforward:
we modify the structure of our Dueling Deep-Q network, 
adding a number of input neurons and output neuron that allow us to encode the new actions, too.
Let us consider a model $\mathcal{M}_{i}^{\mathcal{A}}$ that has been trained using $N$ actions, if a new device is available we enrich the Dueling Deep-Q network with a number $K$ of neurons such that the other $N+M$ actions can be taken into account.
Actually the re-modulation of the Dueling Deep-Q network architecture does not imply automatically the re-training of the model $\mathcal{M}_{i}^{\mathcal{A}}$: if the agent can perform well there in no need to use the new device.
That is the reason why the connections \textit{to} and \textit{from} the new neurons are set to 0: in this way, the trained models can be used in a context of $N + M$ actions, but maintaining the original behavior. 
\subsubsection{DQN-Knowledge-based Adaptation }
This strategy answer the question: what if the other two adaptation  methods fail?
If a novelty that cannot be managed using already available models, is detected, REPTILE should use the data stored in the \textit{Knowledge} for training a new DRL-based agent both if there are new devices and if not.
\subsection{Self-adaptation process}
    
\label{subsec:adaptation}
\begin{figure}
    \centering
    \includegraphics[width = 0.8\linewidth, height = 0.6\linewidth]{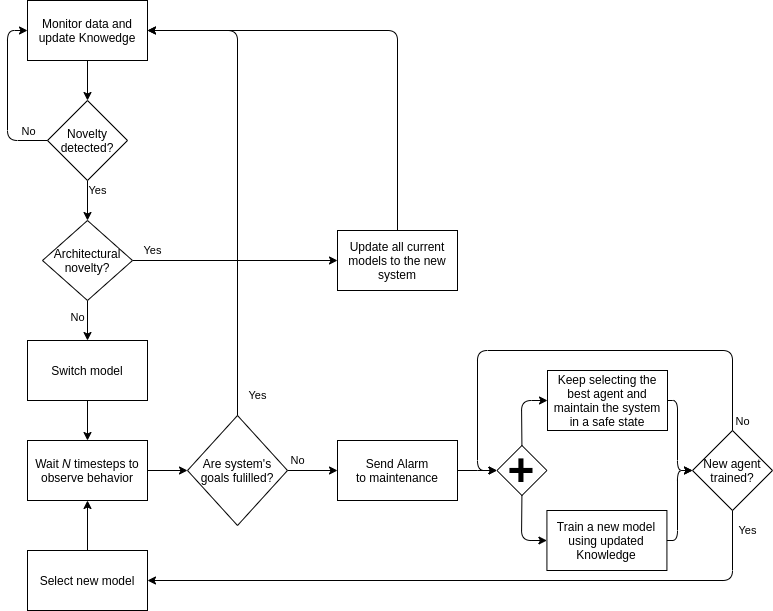}
    \caption{Model Switching Adaptation Diagram.}
    \label{fig:model_switch_diagram}
\end{figure}

The complete self-adaptation process, shown in Figure \ref{fig:model_switch_diagram}, works in the following way: 

\begin{enumerate}
    \item The system continuously monitors the environment and updates the current Knowledge.
    \item The Analyzing module checks for a possible novelty: if an architectural novelty is detected, all current models are updated (neurons are added) and the system goes back to the monitor phase (step 1); on the other hand, if it is not an architectural novelty, the system selects the best model for the current environment, which is the model allowing the RTRL agent to gain the most cumulative reward; if no novelty is detected, the system goes back to the monitor phase.
    \item Once the best model has been selected, the system waits $n$ steps in order to observe the behavior of the agent using the selected model. The quantity $N$ can vary based on the application domain of the SAS.
    \item After $N$ steps, the system checks if the model fulfills all system's goals even in the presence of the novelty: if so, the system has successfully adapted to the novelty and the monitor process starts again (step 1), otherwise an alarm is sent to the maintenance: this is done because it means that the system can not fulfill its goals with its current models, communicating that it has to begin the training phase.
    \item Once the alarm has been fired, the system keeps selecting the best model at its disposal in order to maintain a safe state while a new model is trained using the new environment's data stored in the Knowledge to generate episodes for the RTRL agent.
    \item Once the training process is terminated, the system selects the new model and observes the new behavior (step 3).
\end{enumerate}

\section{Results}
\label{sec:results}

In this paper we applied REPTILE to a HVAC system. Such a system was implemented in Python and it is able to compute the internal temperature of a room by simulating the actions of three devices: a heater, a cooler and a window. The parameters used by the simulator are reported in Table~\ref{tab:simulator_parameters}. Since REPTILE is based on DRL-basd agent we interfaced our simulator with the the Open AI environment that invokes the temperature simulator to calculate the Agents’s next state. 
REPTILE extends the MAPE-K framework with a \emph{Learning} module, based on DRL, and equips the \emph{Knowledge} module with a proactive capability based on the possibility of creating on-the-fly time-varying models of the environment and foreseen their next states. 

\begin{figure}[!h]
    \centering
    \includegraphics[width=6cm,height=8cm,keepaspectratio]{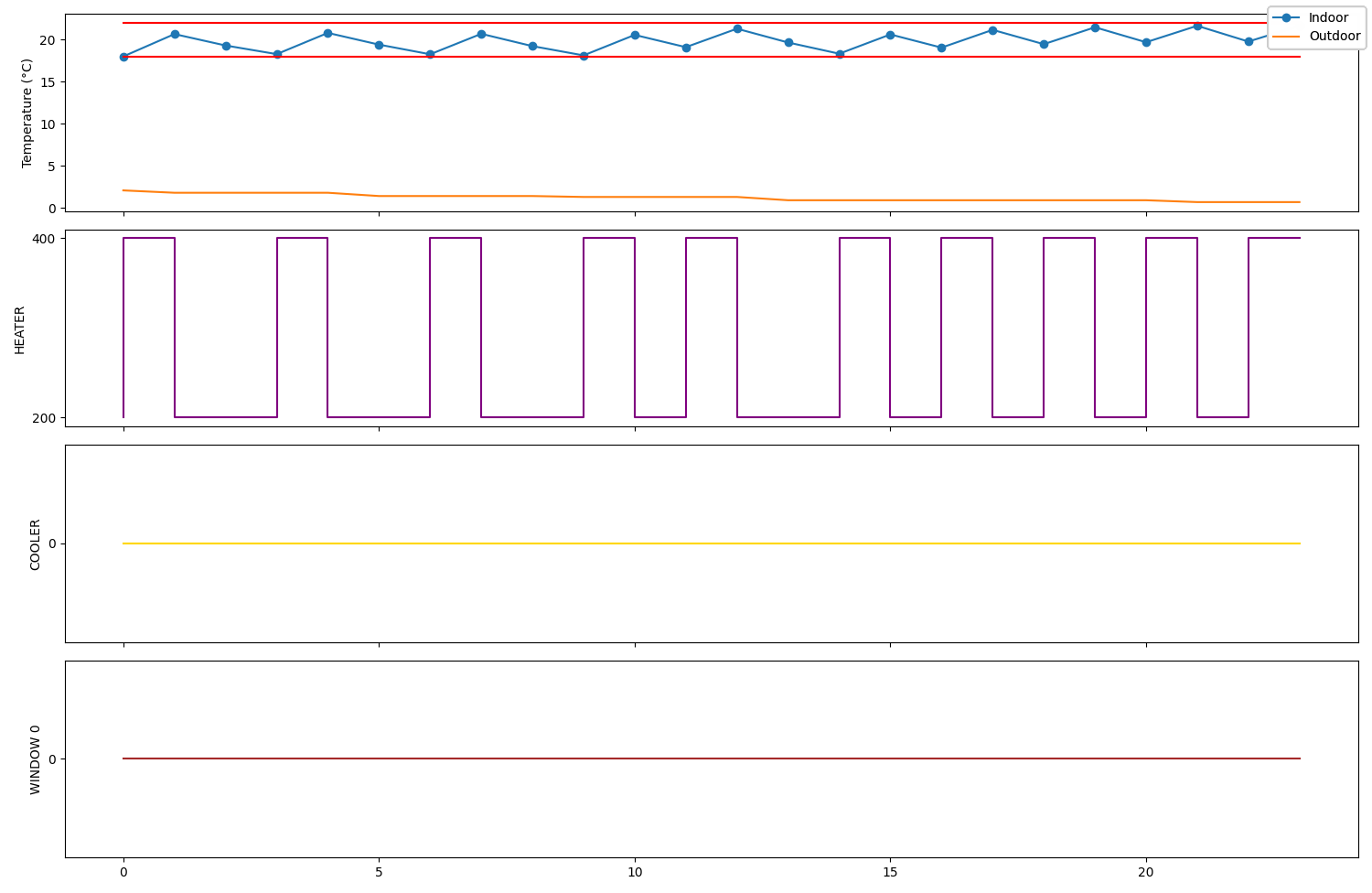} \includegraphics[scale = 0.35]{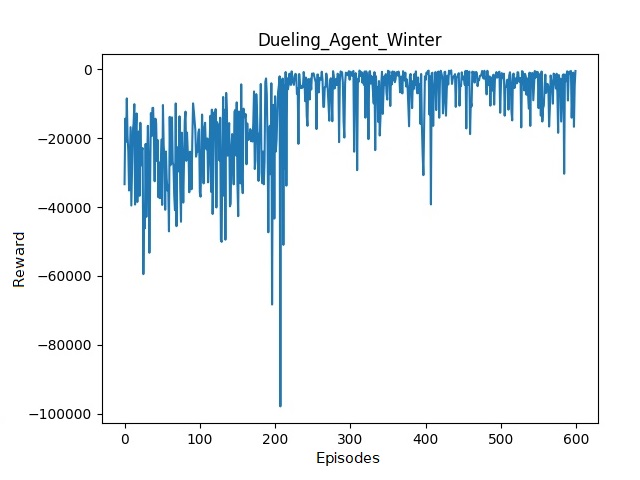}
    \caption{Winter Agent. On the left, the monitoring of the temperature and the actions taken by the agent for maintaining the temperature in the comfort zone; on the right, the mazimization of the reward function is shown}
    \label{fig:dueling_test}
\end{figure}

The \emph{Learning} module comes into play whenever a novelty is detected and, using the knowledge of the environment (or simulate the knowledge from the collected data), it is able to trigger an adaptation strategy: the model switching one, the system architectural one or the re-training one. The \emph{proactive} Knowledge module is able to derive a model of the environment by using the ARIMA formalism and, at the same time, contains some pre-trained models of DRL-based agents, $\mathcal{M}_{1}^{\mathcal{A}}$, $\mathcal{M}_{2}^{\mathcal{A}}$,...,$\mathcal{M}_{k}^{\mathcal{A}}$, and some global models of the changing environment, $\mathcal{M}_{1}^{\mathcal{G}}$,...,$\mathcal{M}_{j}^{\mathcal{G}}$,  in order to avoid to use some specific techniques to derive such models. 
In this work, only the real data about external temperatures, $T_{ext}$, were used. They consist of time series whose values were collected every hour~\cite{beach_dataset,beniaguev_2017}. We pre-processed those data by replicating each value 3 times in order to simulate a sampling frequency of 15 minutes. We also have a model $\mathcal{M}^{\mathcal{G}}$ and a model $\mathcal{M}^{\mathcal{A}}$.
We also have no available new devices.
We show the results in two different operative conditions: summer and winter. 
All the results described below are obtained using a laptop with the Mint Linux OS, 8Gb RAM, a dual-core Intel i7@2.7GHz CPU and a graphic card Nvidia GeForce GTX 940MX.
Initially, we started by having an agent that is trained to behave in winter conditions. We used 600 episodes, with a decreasing $\epsilon$-greedy exploration-exploitation trade-off starting from 1 and an experience reply soft optimization. The learning procedure took about 1 hour to be completed.

The behavior of such an agent is depicted in the left panel of Figure~\ref{fig:dueling_test}: the upper block shows the internal temperature (blue line) remaining inside the comfort zone, $18$\textdegree{}C - $22$\textdegree{}C (red horizontal lines), and the external temperature (orange line).

%

\begin{figure}[!h]
\begin{subfigure}{.48\textwidth}
  \centering

 \hspace*{-0.6cm}\includegraphics[scale = 0.30]{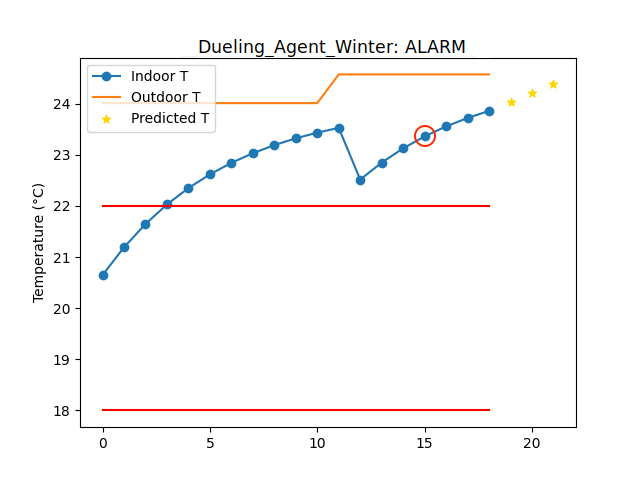}

  \caption{No adaptation}
 \label{fig:alarm}
\end{subfigure}
\begin{subfigure}{.48\textwidth}
\centering
   {%
\stackinset{r}{107pt}{b}{25pt}{\includegraphics[width=60pt]{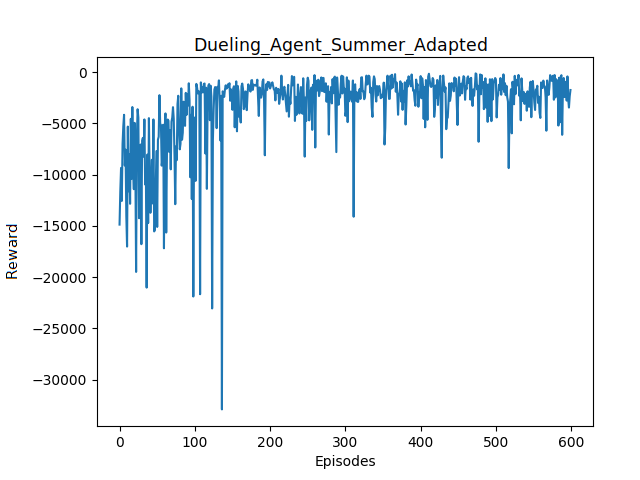}}{%
\includegraphics[scale = 0.30]{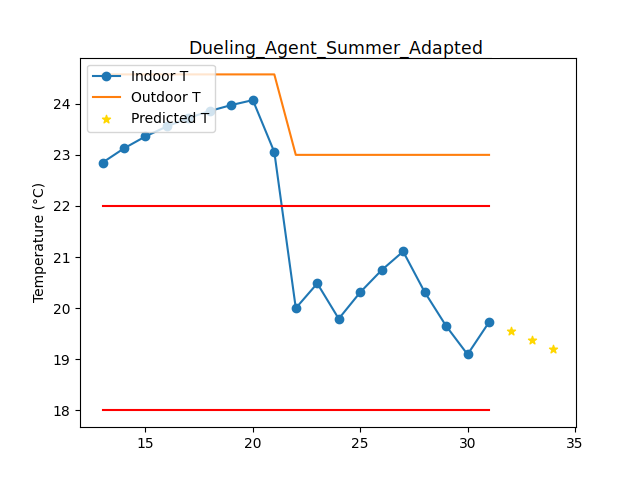}}}
  \caption{Adaptation}
  \label{fig:adaptation_test}
\end{subfigure}
\caption[Test of Novelty Detection.]{Test of Novelty Detection. The Agent can not go back in the Comfort Zone, an alarm is sent and a new model is trained.}
\label{fig:Test_novelty_detection}
\end{figure}


Moreover, the three blocks below display the actions performed by all the tree devices. In particular only the heater keep on increasing and decreasing its power while the cooler and the window do not act.

In order to test the \emph{Model Switching Adaptation} and the \emph{DQN-Knowledge-based Adaptation}, we let the agent behaves in a summer scenario where external temperatures range from 20 to 40 \textdegree{}C. This is a trick for being sure the agent stumbles across a novelty. Figure~\ref{fig:alarm} shows an example of triggering an adaptation process. Horizontal red lines represent our comfort zone while the blue line is the indoor temperature.
After 15 time steps (red circle) the \emph{Analyzing} modules detects, in a \emph{reactive} manner, a violation of our ``soft" constraint: not to be in the Intermediate zone for an hour. Now the system starts the adaptation process finding no new devices (actually the search for new devices is performed at each time step) and only one pre-trained model available: $\mathcal{M}^{\mathcal{A}}$. 
Such a model has been prepared in a way that the system can choose it (its reward is higher) but is not successful in adapting. 
After 3 time steps (remember that the agent perform an action every fifteen minutes), at time step 18, the indoor temperature is again out of range; moreover, the third predicted point (the rightmost yellow point) is in the Danger zone.  At this point, given the impossibility of adapting with the first two strategies, the system chooses to re-train a new DRL-based model using the available $\mathcal{M}^{\mathcal{G}}$ notifying and alarm to the system administrator. The new agent is trained with $\mathcal{M}^{\mathcal{G}}$, containing summer environmental information, for the same 600 episodes as the winter agent.
The inset in Figure~\ref{fig:adaptation_test} shows the maximization of the reward function. The training phase took 1 hour to be completed in this case, too.
Once the new agent has been trained, the system switches form the current model to the new one and it is able to bring the internal temperature back to the Comfort Zone, as depicted in Figure \ref{fig:adaptation_test}.

The above result underlines the ability of the system to adapt its behavior to new operative conditions.
In this case it is worth noticing that the knowledge regarding the new environment (summer scenario) was already in the knowledge module but REPTILE potentially allows to implement any technique to acquire such a knowledge.
Finally, in order to test the \emph{Architectural Adaptation} a new heater with different power values (50 W, 200 W, 250W ,400W) was plugged into the system.

\begin{figure}
    \centering
    \includegraphics[width=7cm,height=10cm,keepaspectratio]{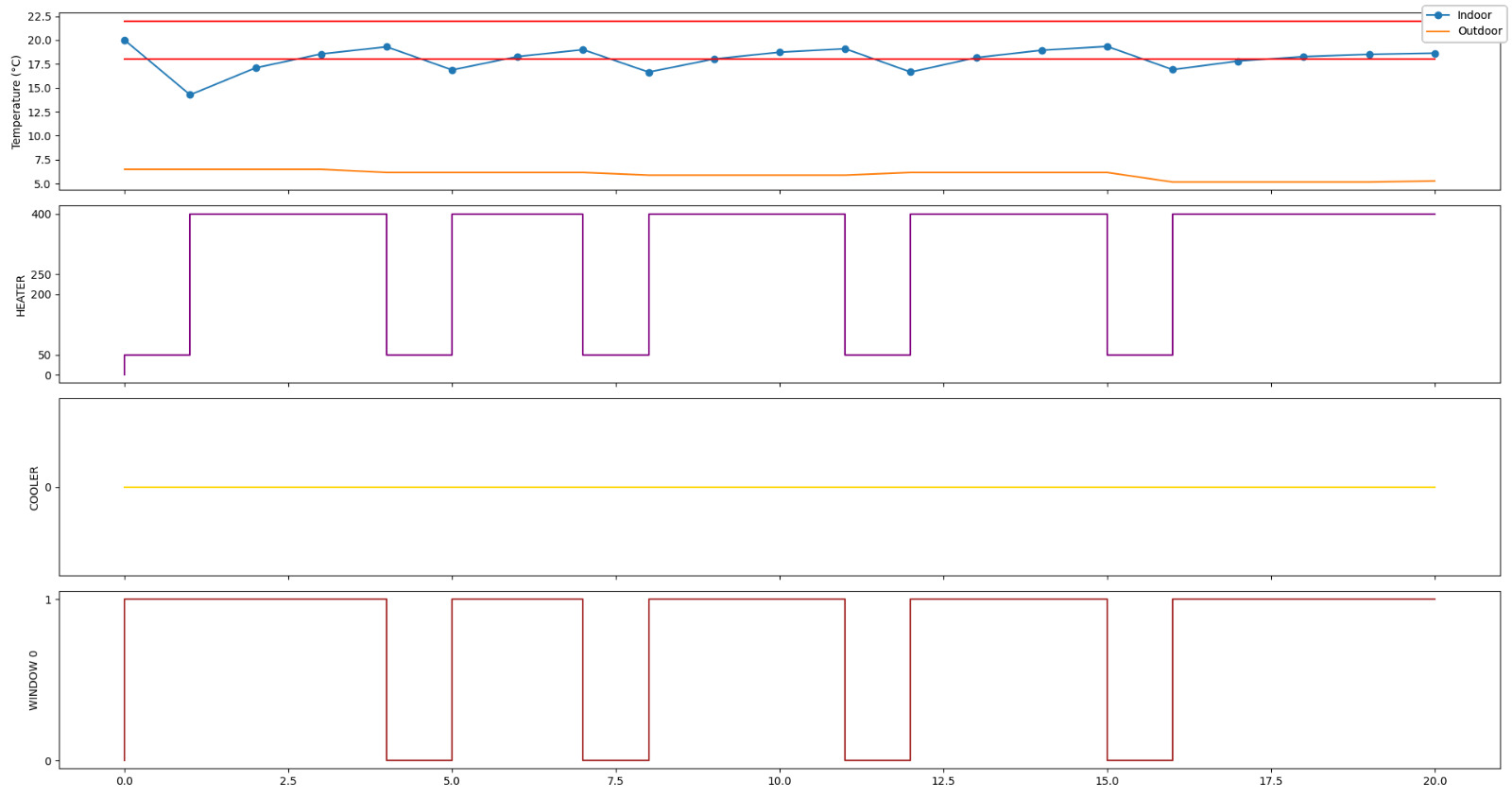}\hspace*{1mm} \includegraphics[scale = 0.35]{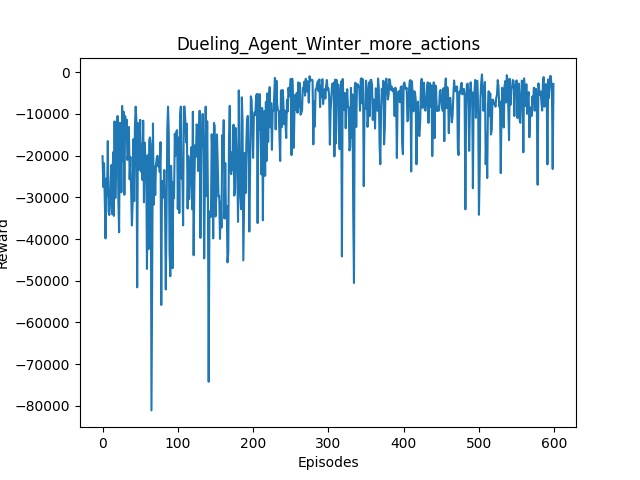}

    \caption{Adaptation after a new device was plugged into the  system. On the left the dynamics of the monitoring of the system with the corresponding actions taken by the DRL-based agent; on the right the maximization of the reward function during the re-training process.}
    \label{fig:change_architecture}
\end{figure}

Once the system detected the new device, it immediately modified the architecture of the Duelling Deep-Q network in order to be able to manage two more actions and then it ran the adaptive procedure: in this the system was running in summer, it was not able to manage the winter external condition so it started the adaptive procedure.
The procedure took again about one hour and produce a behavior like the one shown in Figure~\ref{fig:change_architecture}. Comparing Figure~\ref{fig:change_architecture} (left) with Figure~\ref{fig:dueling_test} it is evident the different actions the agent is performing in order to keep the temperature in the comfort zone.
It is also evident that the agent continues to open and close the window and synchronously raises and lowers the heating. The meaning of this behavior could be that the agent, despite managing to keep the temperature in the comfort zone, would need more episodes to optimize its behavior. The previous hypothesis can be confirmed if we look at Figure~\ref{fig:change_architecture} (right): even though the reward function is maximizing, there are many large amplitude fluctuations. It is worth noticing that for showing the effectiveness of \emph{Architectural Adaptation} the system was induced to apply the \emph{DQN-Knowledge-based Adaptation}, too.

\section{Conclusions}
\label{sec:conclusions}
In this paper, we have presented  REPTILE, a framework for the design of proactive, real-time and self-adaptive systems based on \emph{Deep Reinforcement Learning} (DRL). 
The proposed framework extends IBM’s MAPE-K loop with a \emph{DRL module} that is used to see in advance potential dangerous and unwanted situations, named \emph{novelties}, that may affect the expected behaviour of the considered system. 
Two types of novelties are taken into account in our framework: those related to the \textit{context/environment} and those regarding the \textit{physical architecture} itself~\cite{klos2015adaptive}. 
In the first case, a set of \emph{models}, that are either built via a pre-existing knowledge or via the one acquired \emph{on the fly}, are used to identify the best countermeasures/actions to execute in order to guarantee the satisfaction of some given requirements.   
If no model is available, or when an \emph{architectural novelty} is detected, the \emph{DRL module} is updated to learn new behavioral policies and, consequently, to derive new \emph{models} that will support the adaptation in the new configuration. 
When both these strategies fail, a \emph{DQN-Knowledge-based Adaptation} is executed to re-train the \emph{DRL module}.  

The proposed framework can be applied to both simulated and real/physical systems. However, in the latter case, the appropriate software interfaces should be developed to let REPTILE controls physical devices.
To show an application of REPTILE, in the paper, a \emph{Heating, Ventilation and Air Conditioning system} (HVAC) has been used. The performed experiments confirm the capability of our framework to guide the adaptation procedure. 
We can observe that in the considered example, both the \emph{Model Switching Adaptation} and the \emph{DQN-Knowledge-based Adaptation} seem working well, due to the global model $\mathcal{M}^{\mathcal{G}}$ ussed for training the new \emph{DRL module}. 

As far as we know, this is the first proposal taking into account both the proactivity and the use of real-time deep reinforcement learning paradigm, applied to a Deep Q-Network, for managing data streams.

Even if the \emph{Architectural Adaptation} showed good performances, we believe that further studies are needed to evaluate the possibility of exploiting the \emph{transfer learning paradigm} to avoid the completely re-training of the Deep-Q Network~\cite{kwiatkowski2019task}.

Finally, one of the key point in REPTILE is $\mathcal{M}_{\xi}^{\mathcal{G}}$ that is the model used whenever we need to re-train the DRL-Agent. The better is $\mathcal{M}_{\xi}^{\mathcal{G}}$, the more accurate is $\mathcal{M}_{\xi}^{\mathcal{A}}$. These models can be derived directly from data collected from existing systems by means of different data analysis techniques~\cite{merelli2014rnn,rucco2016survey,piangerelli2020visualising,de2020bayesian,quadrini2021prosps}. In the case of the $\mathcal{M}_{\xi}^{\mathcal{G}}$ models considered in this paper, existing time series have been used.
Nevertheless, further studies are needed to identify the algorithms that are able to derive a model from a (possibly small) set of data, or fast enough to deal with a vary large amount of data.

\bibliographystyle{splncs04}
\bibliography{bibliography}
%




\end{document}